\documentclass[aip,sd,amsmath,amssymb,reprint]{revtex4-1}

\usepackage{graphicx}
\usepackage{dcolumn}
\usepackage{bm}
\usepackage{xcolor}

\makeatother

\begin{document}

\title{Short-term and spike-timing-dependent plasticities facilitate the formation of modular neural networks}

\author{Ewandson L. Lameu}
\affiliation{National Institute for Space Research, S\~ao Jos\'e dos Campos,
S\~ao Paulo, 12227-010, Brazil.}
\email{ewandson.ll@gmail.com}
\author{Fernando S. Borges}
\affiliation{Center for Mathematics, Computation and Cognition, Federal
University of ABC, S\~ao Bernardo do Campo, S\~ao Paulo, 09606-045, Brazil.}
\author{Kelly C. Iarosz}
\affiliation{Institute of Physics, University of S\~ao Paulo, S\~ao Paulo,
05508-900, Brazil.}
\author{Paulo R. Protachevicz}
\affiliation{Program of Post-graduation in Science, State University of Ponta
Grossa, Ponta Grossa, Paran\'a, 84030-900, Brazil.}
\author{Antonio M. Batista}
\affiliation{Institute of Physics, University of S\~ao Paulo, S\~ao Paulo,
05508-900, Brazil.}
\affiliation{Program of Post-graduation in Science, State University of Ponta
Grossa, Ponta Grossa, Paran\'a, 84030-900, Brazil.}
\author{Chris G. Antonopoulos}
\affiliation{Department of Mathematical Sciences, University of Essex, Wivenhoe Park, UK.}
\author{Elbert E. N. Macau}
\affiliation{National Institute for Space Research, S\~ao Jos\'e dos Campos,
S\~ao Paulo, 12227-010, Brazil.}
\affiliation{Federal University of São Paulo, S\~ao Jos\'e dos Campos, S\~ao Paulo, 12247-014, Brazil.}

\date{\today}

\begin{abstract}
The brain has the phenomenal ability to reorganize itself by forming new connections among neurons
and by pruning others. The so-called neural or brain plasticity facilitates the modification of
brain structure and function over different time scales. Plasticity might occur due to external
stimuli received from the environment, during recovery from brain injury, or due to modifications
within the body and brain itself. In this paper, we study the combined effect of
short-term (STP) and spike-timing-dependent plasticities (STDP) on the synaptic strength of
excitatory coupled Hodgkin-Huxley neurons and show that plasticity can facilitate the formation
of modular neural networks with complex topologies that resemble those of networks with
preferential attachment properties. In particular, we use an STDP rule that alters the synaptic
coupling intensity based on time  intervals between spikes of postsynaptic and presynaptic neurons.
Previous works have shown that STDP may induce the appearance of directed connections from high to
low frequency spiking neurons. On the other hand, STP is attributed to the release of neurotransmitters
in the synaptic cleft of neurons that alter its synaptic efficiency. Our results suggest that the
combined effect of STP and STDP with high recovery time facilitates the formation of connections
among neurons with similar spike frequencies only, a kind of preferential attachment. We then pursue
this further and show that, when starting with all-to-all neural configurations, depending on the STP
recovery time and distribution of neural frequencies, modular neural networks can emerge as a direct
result of the combined effect of STP and STDP.
\end{abstract}

\maketitle


\begin{quotation}
Neural or brain plasticity is the remarkable ability of the brain to alter its structure 
and function over time to achieve cognitive functions and perform tasks. Even though, 
plasticity may result as a consequence of external stimuli, it might also occur during 
learning processes and after brain damage. Synaptic plasticity refers to the 
ability of the brain to render synapses weaker or stronger according to neural activity. 
In this paper, we build a neural network model of excitatory coupled Hodgkin-Huxley (HH)
neurons with the ability to modify their synaptic strengths over time, i.e., a neural 
network with short-term (STP) and spike-timing-dependent plasticities (STDP) build in. STDP 
acts on longer time scales compared to STP, with both plasticities playing an important role in brain 
functions. Here, we show that due to STP, neural networks equipped with STDP 
facilitate the formation of synapses among neurons with similar spike frequencies only, a kind
of preferential attachement. Modular neural networks can emerge as a direct result of
the combined effect of STP and STDP, a structure also depicted by real brain networks.
\end{quotation}


\section{INTRODUCTION}

Mathematical models have been employed in neuroscience since early in the 20th century 
\cite{tewari16} to explain experimental findings and elucidate the inner workings of the brain. 
In 1907, Lapicque \cite{lapicque07} proposed the integrate-and-fire model that can 
reproduce the electrical activity of the membrane potential of neurons. It is one of 
the most popular models for studying the dynamic behaviour of neural systems. Later 
in 1952, Hodgkin and Huxley \cite{hodgkin52} explained the ionic mechanisms in the 
cell membrane of neurons and proposed the so-called Hodgkin-Huxley (HH) neural model 
that has, since then, been used extensively to study neural networks \cite{zhu18}.

In particular, neural network models have been used extensively in neuroscience applications, such 
as in studies of neural dynamics \cite{Protacheviczetal2019}, dynamic range 
\cite{Viana14,batista14,Borges2015,Antonopoulos2016}, neural synchronization 
\cite{lameu18a,protachevicz18,Borges2017,Hizanidisetal2016,Antonopoulosetal2019}, 
flow of information \cite{Antonopoulosetal2015,borges18,Antonopoulosetal2016} and 
brain plasticity \cite{romani15,zenke15} to name a few.

Neural plasticity is the ability of the brain to modify its function and 
structure over different time scales \cite{burke06,berlucchi09}. The term was 
initially used by James\cite{james90} in 1890 to propose that phenomena of 
habit in living organisms are due to plasticity.  Cajal \cite{stahnisch02} reported in the early 1900s his research about
regenerative and degeneration changes in brain structure. In 1924, Lashley \cite{lashley24} 
demonstrated experimental evidence of a malleable brain. Konorski \cite{konorski48} 
and Hebb \cite{hebb49} proposed in 1948 and 1949, respectively, that neural 
activities have influence on the connection among neurons. In 1960, Bennet et al. 
\cite{bennett64} carried out experiments with rats and observed chemical and 
anatomical plasticity in the brain. Since then, there have been many theoretical \cite{lameu18b} and 
empirical \cite{rangaraju19} studies aiming to explain and understand brain 
plasticity and its effects in brain structure and function.

In synaptic plasticity, the synapses among neurons are potentiated or depressed 
in time according to the activity of the neurons \cite{abbott00,borges17a}. 
Recent works \cite{mcdonnell17,tass18} have shown that short-term (STP) and spike-timing-dependent 
plasticities (STDP) are different forms of neural processes leading to synaptic 
modifications. In particular, STDP depends on the relative 
timing of presynaptic and postsynaptic neural spikes \cite{markram12,borges16}. 
This type of plasticity can lead to various dynamical phenomena and coupling structures,
such as stable localized structures \cite{clopath10}, stimulation-induced synchronization or
desynchronization \cite{tass12}, noise-enhanced synchronization \cite{popovych13,lucken16}, and nontrivial topology \cite{borges17b}. 
The STDP mechanism plays a role in temporal coding of information by spikes \cite{clopath10, gestner96}.
On the other hand, 
STP is attributed to the release of neurotransmitters in the synaptic cleft of 
neurons that alter its synaptic efficiency and acts on shorter time scales, ranging 
from milliseconds to hundreds to thousands of milliseconds \cite{stevens95,abbott97,zucker02,hennig13}. 
As in the case with STDP, STP can have a great influence on the network's dynamical
behavior. For instance, it may stabilize the parametric working memory \cite{itskov11}, contribute
to the emergence of  spontaneous traveling waves \cite{york09}, or 
induce phase changes in neural postsynaptic spiking \cite{mcdonnell17}.

Here, we extend the work in Borges et al. \cite{borges17b} which was focused on 
STDP only and study the combined effect of STP and STDP in neural networks of 
excitatory coupled HH neurons. The plasticity terms that 
model STDP in the equations in Sec. \ref{HH_network_STP_STDP} are based on 
the experimental results by Bi and Poo\cite{bi98,bi01} that were performed on 
excitatory synapses and on theoretical results by Abbott et al. \cite{abbott97} and
Popovych et al. \cite{popovych13} (STDP). The results in Bi and Poo\cite{bi98,bi01} 
show that STDP is a function of the relative timing of 
postsynaptic and presynaptic spikes and is theoretically backed by
the Hebbian synaptic learning rule \cite{hebb49}. Instead, STP depends on the 
neural recovery dynamics \cite{liu03,zucker89}. McDonnell and Graham 
\cite{mcdonnell17} used mathematical analysis and numerical simulations to 
show that STP induces phase changes in neural postsynaptic spiking.
In our work, we start by studying the simplest case of a pair of HH neurons 
for a range of spike frequencies, aiming to understand how connectivity between them  
changes by the combined effect of STP and STDP. Next, we build an initially, all-to-all (globally) connected 
network of HH neurons and consider the simultaneous effect of STP and STDP in a range of coupling strengths. We 
show that STP plays an important role in topology changes in neural networks with STDP.  
Indeed, we find that for high STP recovery time, only neurons with similar spike frequencies tend to connect,
a form of preferential attachement. More importantly, our results show that, when starting 
with all-to-all networks, depending on the STP recovery time and distribution of neural 
frequencies, modular neural networks can emerge as a direct result of the combined 
effect of STP and STDP, a structure depicted by neurophysiological and 
experimental studies \cite{bifone16,betzel16}. For the considered setup, STP plays a
balancing role: while STDP tends to synchronize all neurons in one cluster, the STP destroys
the strong synchronization and leads to a modular structure.

The paper is structured as follows: In Section II, we introduce the general mathematical 
model of HH neural networks with STP and STDP and, in Section III, we present our analysis 
and results based on numerical simulations, that show the effects of both plasticities, initially 
on a pair of neurons and then, on a network of 100 HH neurons. Finally, we present the 
conclusions of our work in the last section.


\section{A HODGKIN-HUXLEY NEURAL NETWORK with STP and STDP}\label{HH_network_STP_STDP}

We buid use a neural network model of $N$ HH neurons coupled
with excitatory chemical synapses, equipped with STP and STDP rules based on the experimental
results by Bi and Poo\cite{bi98,bi01} and theoretical 
models proposed by Popovych et al. \cite{popovych13} (STDP) and McDonnell and Graham 
\cite{abbott97, mcdonnell17,liu03}(STP).

Specifically, the HH neural network model considered is given by
\begin{eqnarray}
C\dot{V_i} &=& I_i-g_{\rm K}n_i^{4}(V_i-E_{\rm K})-
g_{\rm Na}m_i^{3}h_i(V_i-E_{\rm Na}) \label{voltage_eq} \nonumber \\
&-&g_{L}(V_i-E_{\rm L})+(V_r-V_i)\sum_{j=1}^N\varepsilon_{ij} f_jD_j \label{HH},\\
\dot{n}_i &=& \alpha_{n_i}(V_i)(1-n_i)-\beta_{n_i}(V_i)n_i,\\
\dot{m}_i &=& \alpha_{m_i}(V_i)(1-m_i)-\beta_{m_i}(V_i)m_i,\\
\dot{h}_i &=& \alpha_{h_i}(V_i)(1-h_i)-\beta_{h_i}(V_i)h_i,\\
\dot{f}_i &=& -\frac{f_{i}}{\tau_s}, \label{f} \\
\dot{D}_i &=& \frac{1-D_i}{\tau_D}, \label{D}
\end{eqnarray}
where $C$ ($\mu$F/cm$^2$) is the membrane capacitance and $V_i$ (mV) the 
membrane potential of neuron $i$ at time $t$ (where $i=1,\dots,N$). 
$I_i$ ($\mu$A/cm$^2$) is the constant current density of neuron $i$ and $\varepsilon_{ij}$
represents the matrix of coupling weights between neurons 
$i$ and $j$. $n_i$ and $m_i$ are the activation of potassium and 
sodium functions, respectively, and $h_i$ the inactivation of sodium 
function. Parameters $g$ and $E$ are associated with the conductance and 
reversal potential of each ion, respectively, and $V_r$ is the excitatory 
reversal potential.

The various rate functions in Eqs. \eqref{voltage_eq} - \eqref{D} are given by
\begin{eqnarray}
\alpha_{n}(v) &=& \frac{0.01v+0.55}{1-\exp\left(-0.1v-5.5\right)},\\
\beta_{n}(v) &=& 0.125\exp\left(\frac{-v-65}{80}\right),\\
\alpha_{m}(v) &=& \frac{0.1v+4}{1-\exp\left(-0.1v-4\right)},\\
\beta_{m}(v) &=& 4\exp\left(\frac{-v-65}{18}\right),\\
\alpha_{h}(v) &=& 0.07\exp\left(\frac{-v-65}{20}\right),\\
\beta_{h}(v) &=& \frac{1}{1+\exp\left(-0.1v-3.5\right)},
\end{eqnarray}
where $v=V/[{\rm mV}]$ is the membrane voltage $V$ of Eq. \eqref{voltage_eq} in 
millivolts (mV) divided by 1mV to render the variable dimensionless. We consider $C=1\mu$F/cm$^{2}$, 
$g_{\rm K}=36$mS/cm$^{2}$, $g_{\rm Na}=120$mS/cm$^{2}$, $g_{\rm L}=0.3$mS/cm$^{2}$, 
$E_{\rm K}=-77$mV, $E_{\rm Na}=50$mV, $E_{\rm L}=-54.4$mV and $V_r=20$mV. 

\begin{figure}[htbp]
\begin{center}
\centering\includegraphics[width=0.4\textwidth]{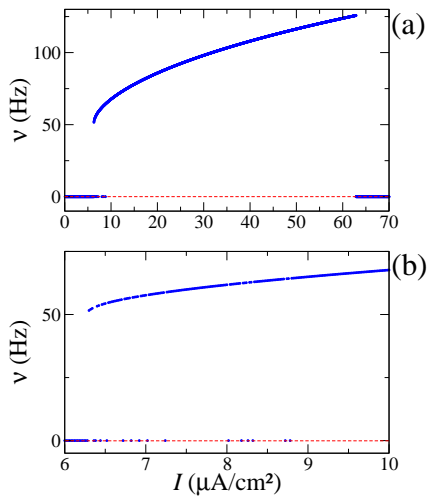}
\caption{Natural frequency $\nu$ as a function of the current $I$ for
a single HH neuron. Panel (b) is a zoom-in of panel (a) for currents $I$ in $[6,10]\mu$A/cm$^2$.}
\label{fig1}
\end{center}
\end{figure}

Figure \ref{fig1}(a) shows the spike (or natural) frequency $\nu$ (Hz) of a single HH neuron 
as a function of current $I$. The spikes were numerically computed when the voltage $V$  
crosses the threshold of 0mV, increasing from negative to positive values. In the simulations 
of 100 coupled HH neurons in Sec. \ref{STP_STDP_HH_networks}, we consider $I_i$ randomly distributed in the interval 
$[10,30]\mu$A/cm$^2$, leading to spike frequencies $\nu_i$ in the interval $[70,100]$Hz. 
This interval avoids the regime observed for $I$ in $[6,10]\mu$A/cm$^2$, as shown in 
Fig. \ref{fig1}(b), where it is clear that for some $I$ values, the neuron does not spike as
its natural frequency $\nu$ is zero. This choice of interval also allows for
the monotonic increase of the natural frequencies $\nu_i$ without reaching currents $I$
bigger than $60\mu$A/cm$^2$ that correspond to a non-spike regime.

In Eq. \eqref{f}, $f_i$ is the strength of the effective synaptic output 
current from neuron $i$ to neuron $j$ and $\tau_s$ (ms) the synaptic time constant, 
fixed at $\tau_s=2.728$ms. When neuron $i$ spikes, $f_i$ is updated ($f_i\to 1$) before it starts to decay exponentially.  

Equation \eqref{D} models STP \cite{liu03,mcdonnell17} with $\tau_D$ (ms) being the 
recovery time constant, related to biological mechanisms such as the depletion 
of release-ready neurotransmitter vesicles at the presynaptic terminal 
\cite{zucker02,zucker89}. We assume that every time neuron $i$ spikes, 
the update rule $D_i\to (D_i-d)$ is applied. Biologically, $D_i$ could represent 
the vesicles that can be used to transmit a signal from the presynaptic to the
postsynaptic neuron. Therefore, the update rule means that the amount $D_i$ 
of available vesicles is decreased by $d=0.1$ in every spike of neuron $i$,
and then it recovers according to Eq. \eqref{D}. In that framework, $D_i$ lies in $[0,1]$
as if it happens $D_i$ to be negative it means that the neuron used all the stored vesicles
and when this happens, $D_i$ is reset to 0. On the other hand, when $D_i$ is equal
to 1, it means that all neurotransmitter vesicles are restored.  

\begin{figure*}[htbp]
\begin{center}
\centering\includegraphics[width=0.9\textwidth]{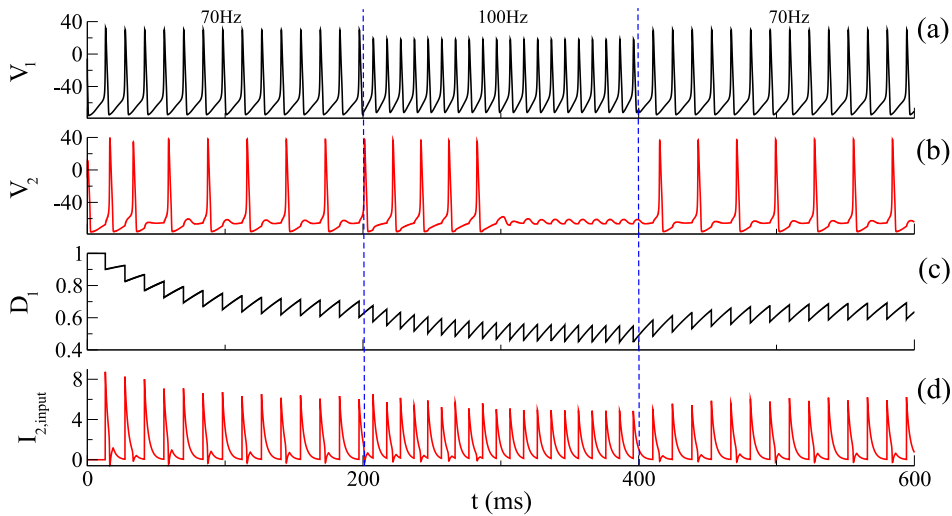}
\caption{The effect of STP on a pair of unidirectionally connected HH neurons, 
where neuron 1 is connected to neuron 2 but not vice versa: Temporal evolution of (a) $\nu_1$, 
(b) $\nu_2$, (c) $D_1$ and (d) $I_{2,\rm input}=(V_r-\nu_2)\varepsilon_{21} f_1D_1$ 
with $\varepsilon_{12}=0$, $\varepsilon_{21}=0.1$, $I_2=0$ and $\tau_D=50$ms. Note that, for
$\nu_1=70$Hz and $\nu_1=100$Hz, we used $I_1=10.97\mu$A/cm$^2$ and $I_1=31.8\mu$A/cm$^2$, respectively.}
\label{fig2}
\end{center}
\end{figure*}

Figure \ref{fig2} shows the effect of STP on a pair of neurons coupled with a 
unidirectional connection from neuron 1 to neuron 2. This is implemented by 
fixing $\varepsilon_{12}=0$ (i.e. the connectivity strength from neuron 2 
to neuron 1 is 0) and $\varepsilon_{21}=0.1$ for the connectivity strength from 
neuron 1 to 2. In this study, we have set the STP recovery time $\tau_D$ at 
$50$ms and $I_2$ at $0$, so that neuron 2 spikes only when it receives a strong 
enough input $I_{2,\rm input}=(V_r-V_2)\varepsilon_{21} f_1D_1$ from neuron 1. 
When the spiking frequency $\nu_1$ in Fig. \ref{fig2}(a) changes from $70$Hz to 
$100$Hz, $\nu_2$ in Fig. \ref{fig2}(b) exhibits a delayed alteration in its 
dynamic behaviour as its amplitude plummets at about 300ms. We 
appreciate further this phenomenon through the temporal evolution of $D_1$ in 
Fig. \ref{fig2}(c) and the input current $I_{2,\rm input}$ 
received by neuron 2 in Fig. \ref{fig2}(d). Comparing the 70Hz-regime 
with the 100Hz-regime in Figs. \ref{fig2}(c) and (d), one can see that $D_1$ 
decreases with the increase of the spike-frequency of neuron 1 to $100$Hz, and 
consequently $I_{2,\rm input}$ becomes less intense as it is not strong enough 
to cause spikes in the activity of neuron 2. When neuron 1 returns to $70$Hz spike frequency, there is more 
time for $D_1$ to recover, thereby to increase the intensity of $I_{2,\rm input}$ 
which triggers again spikes in neuron 2. Generalizing this, one might 
say that STP makes neurons more sensitive to spike frequency changes.

\begin{figure}[htbp]
\begin{center}
\centering\includegraphics[width=0.4\textwidth]{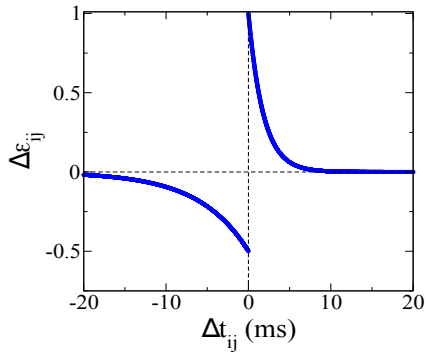}
\caption{Plot of $\Delta \varepsilon_{ij}$ that models STDP as a function 
of the difference between the spike times of the postsynaptic neuron $i$ and presynaptic 
neuron $j$. Note the discontinuity at $\Delta t_{ij}=0$ms.}
\label{fig3}
\end{center}
\end{figure}

Moving now to the other form of plasticity introduced to the model, STDP 
gives rise to changes in the synaptic strength by means of the update function
\cite{bi98,popovych13}
\begin{equation}
\varepsilon_{ij}\to\varepsilon_{ij}+10^{-3}\Delta\varepsilon_{ij},
\end{equation}
where
\begin{equation}
\Delta \varepsilon_{ij}= \left\{
\begin{matrix}
\displaystyle \varepsilon_+=A_{1}e^{(-\Delta t_{ij}/\tau_{1})}, & {\rm if} &
\Delta t_{ij}>0 \\
\displaystyle \varepsilon_- =-A_{2}e^{({\Delta t_{ij}/\tau_{2}})}, & {\rm if} &
\Delta t_{ij}<0 \\
\displaystyle 0, & {\rm if} & \Delta t_{ij}=0  \label{STDP}
\end{matrix} 
\right. .
\end{equation}
Here, $\Delta t_{ij}=t_i-t_j$ is the difference between the spike times of 
the postsynaptic ($t_i$) and presynaptic ($t_j$) neurons $i$ and $j$, 
respectively. Figure \ref{fig3} shows the plot of the plasticity function 
$\Delta \varepsilon_{ij}$ calculated from Eq. \eqref{STDP} for $A_{1}=1$, $A_{2}=0.5$, $\tau_{1}=1.8$ms, $\tau_{2}=6$ms,
and $\Delta t_{ij}$ varying from $-20$ms to $20$ms. This update rule is applied
everytime the postsynaptic neuron $i$ spikes.

To understand better the effects of the recovery time $\tau_D$, we consider in
the next the case of two, unidirectionally connected neurons where neuron 1
(presynaptic) is connected to neuron 2 (postsynaptic), but not vice versa. We
evaluate how the average input current ${\bar I}_{2,\rm{input}}$ and amount of
information exchanged change with the increase of the recovery time $\tau_D$ by
comparing cases where the neurons have similar and dissimilar spike frequencies.

To quantify the exchange of information between the 2 neurons, we compute the Mutual
Information (MI) \cite{Shannon, Kullback, Dobrushin}. MI 
can be understood as the amount of dependence (or uncertainty) between two random
variables $X$ and $Y$, and is given by
\begin{eqnarray}
\mbox{MI}_{XY}(n) &=& \sum_{i}^{n}\sum_{j}^{n}P_{XY}(i,j){\rm log}\left ( \frac{P_{XY}(i,j)}{P_X(i)P_Y(j)} \right ) \label{MI}
\end{eqnarray}
where the probability of a random event $i$($j$) to occur in $X$($Y$) is given
by $P_X(i)$($P_Y(j)$). The joint probability $P_{X,Y}(i,j)$ gives the probability
of $i$ to occur in $X$ and $j$ in $Y$, simultaneously. In this context, the number
of random events in $X$ and $Y$ is denoted by $n$. 

\begin{figure}[htbp]
\begin{center}
\centering\includegraphics[width=0.5\textwidth]{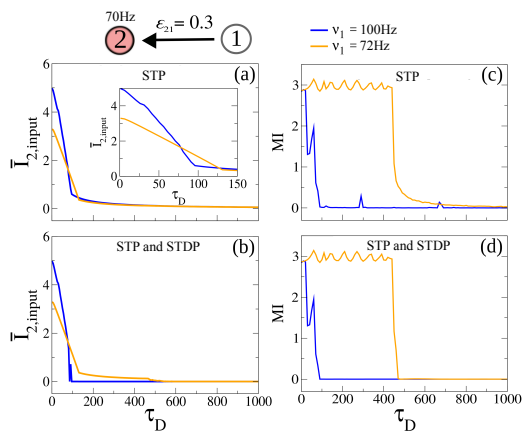}
\caption{The average input current ${\bar I}_{2,\rm{input}}$ neuron 2 receives from neuron 1
and mutual information, MI, as a function of the recovery
time $\tau_D$ for the following cases: (a), (c) only STP and (b), (d) STP and STDP. 
Neuron 2 has a fixed spike frequency $\nu_2=70$Hz ($I_2=10.97\mu$A/cm$^2$). The blue curve
represents the case where $\nu_1=100$Hz ($I_1=31.8\mu$A/cm$^2$) and the orange curve, the case where
$\nu_1=72$Hz ($I_1=11.88\mu$A/cm$^2$). We use $\varepsilon_{21}=0.3$
and $\varepsilon_{12}=0$, implying that only $\varepsilon_{21}$ changes when STDP is considered. 
The simulations run for $200\times10^3$ms, ${\bar I}_{2,\rm{input}}$
is calculated over the last $10\times10^3$ms and, the time-series for $\nu_1$ and $\nu_2$ used
for the calculation of MI are recorded from the last $100\times10^3$ms of the simulations.}
\label{fig4}
\end{center}
\end{figure}

In the next, we consider the cases where the system has only STP
(see Fig. \ref{fig4}(a), (c)) and, STP and STDP (see Fig. \ref{fig4}(b), (d)).
In both cases, we consider $\varepsilon_{21}=0.3$ and $\varepsilon_{12}=0$,
but for the case of STD and STDP (Fig. \ref{fig4}(b), (d)), STDP acts only on
$\varepsilon_{21}$ (unidirectional connection) and ensures that the spike frequency
of presynaptic neuron 1 remains unchanged. The spike frequency of neuron 2 was fixed
at $\nu_2=70$Hz ($I_2=10.97\mu$A/cm$^2$) and, for the blue curves in Fig. \ref{fig4},
neuron 1 has $\nu_1=100$Hz ($I_1=31.8\mu$A/cm$^2$) and for the orange curves, $\nu_1=72$Hz
($I_1=11.88\mu$A/cm$^2$). We use as random variables to calculate MI, the time series of
the voltage of both neurons, namely $X=V_1$ and $Y=V_2$. We let the system evolve
for $200\times10^3$ms for each $\tau_D$ and used the last $100\times10^3$ms to calculate MI.
For the calculation of ${\bar I}_{2,\rm{input}}$, we consider the last $10\times10^3$ms
of the simulations.

Observing Fig. \ref{fig4}(a) (STP), one can see that for recovery times $\tau_D<75$ms, the
average current ${\bar I}_{2,\rm{input}}$ for $\nu_1=100$Hz (blue curve) is bigger than
${\bar I}_{2,\rm{input}}$ for $\nu_1=72$Hz (orange curve) (see also the inset in
Fig. \ref{fig4}(a)). Interestingly, this change in the region $75$ms$<\tau_D<125$ms
where ${\bar I}_{2,\rm{input}}$ for $\nu_1=72$Hz (orange curve) is bigger than
${\bar I}_{2,\rm{input}}$ for $\nu_1=100$Hz (blue curve). For $\tau_D>125$ms, both
${\bar I}_{2,\rm{input}}$ curves assume similar values and settle asymptotically to
zero with further increasing in $\tau_D$.

In Fig. \ref{fig4}(b) (STP and STDP), we see that the action of STP and STDP causes
${\bar I}_{2,\rm{input}}$ for $\nu_1=100$Hz to drop to 0 at $\tau_D\approx100$ms whereas
${\bar I}_{2,\rm{input}}$ for $\nu_1=72$Hz (orange curve) stays positive until
$\tau_D\approx 480$ms. Panels (c) and (d) show clearly the impact of STP and STDP to MI.
In particular, MI fluctuates around 3 for $\tau_D<480$ms in both panels. However, both
panels show that as $\tau_D$ increases from 0ms, MI for $\nu_1=100$Hz (blue curve) approaches
zero at $\tau_D\approx 100$ms, while for $\nu_1=72$Hz, it remains close to 3 until
$\tau_D\approx 480$ms. These results show that STP make the influence of slower (or close
frequency) neurons to become greater than that of faster (or very different frequency)
neurons as $\tau_D$ increases. In conjunction with STDP, this effect occurs due to the
decrease of ${\bar I}_{2,\rm{input}}$ that makes the firing times uncorrelated (less
synchronized), which then causes the coupling to disappear, i.e. $\varepsilon_{21}\rightarrow 0$.
This decoupling process tends to occur for smaller $\tau_D$ values as the difference between
$\nu_1$ and $\nu_2$ is amplified. We address this further later when we discuss the results in Fig. \ref{fig5}.

\section{EFFECTS OF STP ON HH NEURAL NETWORKS WITH STDP}\label{STP_STDP_HH_networks}

Neural networks with STDP and random synaptic input were studied by
Popovych et al. \cite{popovych13}. The authors reported that the mean 
synaptic coupling depends on the noise intensity. Recently, the authors in
\cite{borges17b} showed that STDP induces non-trivial topology in 
neural networks. Here, we extend this work and build a neural network of 
$N=100$ HH neurons to study the combined effect of STP and 
STDP on the structure of the network and in particular, on its connectivity.
\begin{figure*}[htbp]
\begin{center}
\centering\includegraphics[width=0.9\textwidth]{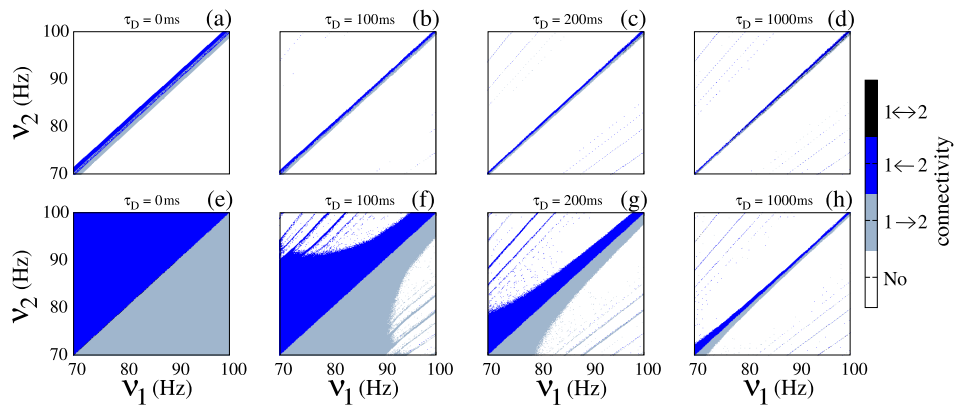}
\caption{Parameter spaces $\nu_1\times\nu_2$ for a pair of initially 
uncoupled neurons for recovery times (a) $\tau_D=0$ms, (b) $\tau_D=100$ms, (c) $\tau_D=200$ms 
and (d) $\tau_D=1000$ms, and for a pair of initially coupled neurons for 
(e) $\tau_D=0$ms, (f) $\tau_D=100$ms, (g) $\tau_D=200$ms and (h) $\tau_D=1000$ms. 
Note that $\nu_1$ and $\nu_2$ vary in $[70,100]$Hz and that the colour bar shows 
the direction of synaptic connectivity, where white accounts for the 
uncoupled case (denoted ``No''), grey for the case where neuron 1 is 
connected to neuron 2 (i.e. $1\rightarrow 2$), blue for the case where neuron
2 is connected to neuron 1 (i.e. $1\leftarrow 2$) and black for the 
bidirectional connection (i.e. $1\leftrightarrow 2$).}
\label{fig5}
\end{center}
\end{figure*}

We start by analysing neural connectivity under the effect of STP and STDP. 
To this end, we consider a pair of HH neurons with STP and STDP, with the 
coupling strengths $\varepsilon_{12}$ and $\varepsilon_{21}$ varying in 
$[0,0.3]$, and the connectivity threshold set at $0.01$.
This threshold is chosen because the coupling weights that should vanish in 
time actually oscillate around 0 assuming very small values as STDP is always 
present and thus, affecting them. As a consequence, the time averages of the coupling 
weights are not 0 but very close to 0. Moreover, for coupling weights smaller 
than this threshold, we notice that neurons influence each other only slightly.

Initially, the pair of neurons is uncoupled (i.e. $\varepsilon_{12}=\varepsilon_{21}=0$) 
or bidirectionally coupled with $\varepsilon_{12}=\varepsilon_{21}=0.3$. 
Figure \ref{fig5} shows the direction of connectivity, after a transient 
time, for different natural frequencies $\nu_1$ and $\nu_2$ in $[70, 100]$Hz. The 
direction is coloured according to the connection from the output of one neuron to the 
input to the other neuron (directions of arrows in the colour bar): white accounts for the 
uncoupled case denoted in the colour bar as ``No'' ($\bar\varepsilon_{ij}<0.01$), 
grey for the case where neuron 1 is connected to neuron 2 (i.e. $1\rightarrow 2$), 
blue for the case where neuron 2 is connected to neuron 1 (i.e. $1\leftarrow 2$), 
and black for the bidirectional connection (i.e. $1\leftrightarrow 2$). For 
initially uncoupled neurons, the connections remain only for neurons with very 
similar frequencies, from those with faster spike frequencies to 
those with slower as depicted in Fig. \ref{fig5}(a) ($\tau_D=0$ms), \ref{fig5}(b) 
($\tau_D=100$ms), \ref{fig5}(c) ($\tau_D=200$ms), and \ref{fig5}(d) ($\tau_D=1000$ms). 
We note the absence of bidirectional connections and that the increase of the recovery time $\tau_D$ 
gives rise to a narrower region of directed connections. With regard to neurons 
starting with bidirectional connections, we observe that for $\tau_D=0$ms, neurons 
still remain unidirectionally connected (for all natural frequencies) from the 
high to the low frequency neurons (see Fig. \ref{fig5}(e)). Increasing $\tau_D$ 
to $100$ms (see Fig. \ref{fig5}(f)), $200$ms (Fig. \ref{fig5}(g)) and $1000$ms 
(Fig. \ref{fig5}(h)), we observe a decrease in the area that represents 
connectivity, implying that the region of high spike frequencies is more affected 
by the influence of STP. Therefore, STDP makes the connections increase from 
faster spiking to slower spiking neurons and STP decreases the influence of 
hight frequency neurons, allowing connections only for those neurons with 
similar spike frequencies.

\begin{figure}[htbp]
\begin{center}
\centering\includegraphics[width=0.49\textwidth]{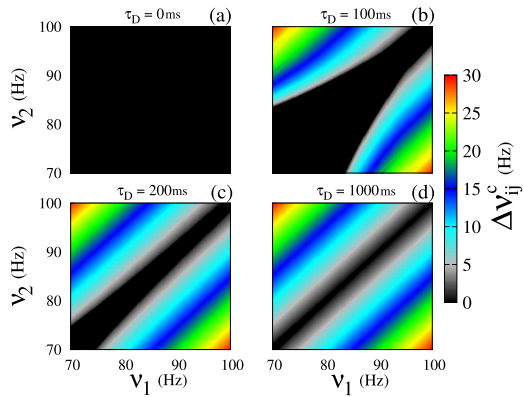}
\caption{Parameter space $\nu_1\times\nu_2$ for increasing values of 
$\tau_D$: (a) for $\tau_D=0$ms, (b) for $\tau_D=100$ms, (c) for $\tau_D=200$ms 
and (d) for $\tau_D=1000$ms. Note that $\nu_1$ and $\nu_2$ vary in $[70,100]$Hz 
and the colour bar represents $\Delta\nu^c_{ij}$ values.}
\label{fig6}
\end{center}
\end{figure}

The reason for the initially, bidirectionally coupled system to have larger 
areas of connectivity is related to the difference in frequency 
$\Delta\nu^c_{ij}$ of neurons when coupled (or the synchronization level). 
We note that this difference is not equal to the difference between their 
natural frequencies $\nu_i - \nu_j$, i.e. $\Delta\nu^c_{ij} \neq \nu_i - \nu_j$. 
In Fig. \ref{fig6}, we present the calculation of the average $\Delta\nu^c_{ij}$ 
(colour bar) for the first $t=4000$ms for the system with STP only. We 
consider $N=2$, $\varepsilon_{12}=\varepsilon_{21}=0.3$ and vary $\nu_1$ 
and $\nu_2$ in $[70,100]$Hz. Figure \ref{fig6}(a) shows the result for 
$\tau_D=0$ms where, despite the natural frequencies of the neurons, $\Delta\nu^c_{ij}$ is
approximately equal to 0 (black region). When comparing these results with
the results in Fig. \ref{fig5}(e), once can see that 
this corresponds to a connected region. For increasingly bigger recovery 
times $\tau_D$ (see Fig. \ref{fig6}(b) for $\tau_D=100$ms, \ref{fig6}(c) for 
$\tau_D=200$ms and \ref{fig6}(d) for $\tau_D=1000$ms), one can observe 
a decrease in the size of the black area, similar to the size of the area seen 
in Fig. \ref{fig5}(f), (g) and (h). Thus, one can infer that what defines 
a connected configuration in the case of STDP is the difference in spike
frequency among neurons or how synchronized they become. The non-homogeneous 
distribution of connected areas in relation to the main diagonal in 
Figs. \ref{fig5} occurs due to the non-linear variation of neural 
frequencies in relation to the received external current. In Fig. \ref{fig1}, 
we see that for low currents (i.e. $I\in[10,25])$, a small increase in $I$ 
can cause a bigger variation in the frequencies when compared to the interval 
where $I>30$. Therefore, low frequency neurons are more sensitive to changes 
in their external currents, which facilitates their synchronization with 
neurons with similar frequencies. The more synchronized they become, the 
more they remain connected. Coming back to the effect of STP, we 
find that it decreases the influence of the faster neurons on the slower 
ones, and consequently leads to the increase of their frequency differences 
and to the suppress of their synchronization. That then leads to the depression
of the connectivity via STDP.

\begin{figure*}[htbp]
\begin{center}
\centering\includegraphics[width=0.9\textwidth]{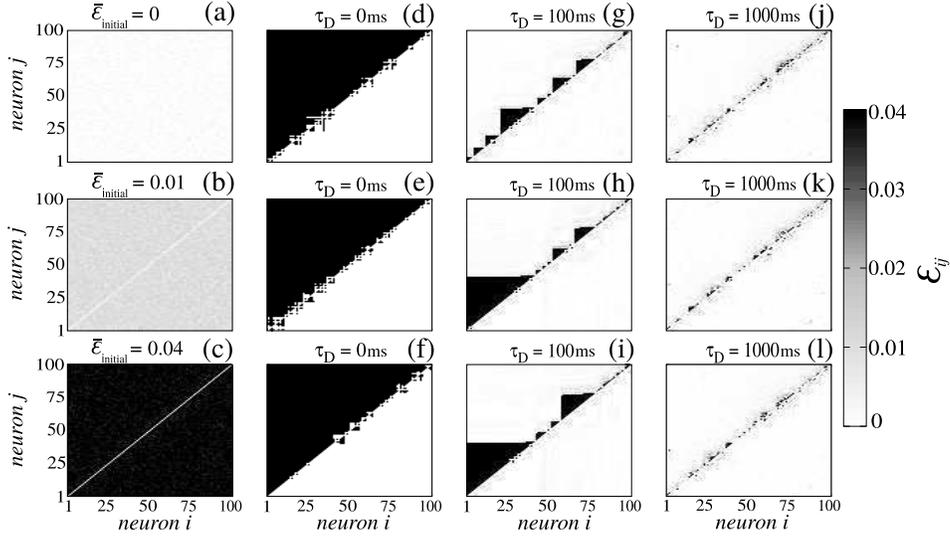}
\caption{The combined effect of STP and STDP on connectivity in networks of $N=100$ excitatory coupled HH
neurons and emergence of modular neural networks. The initial couplings used are: In (a) $\bar\varepsilon_{\rm initial}=0$, in (b) 
$\bar\varepsilon_{\rm initial}=0.01$ and in (c) $\bar\varepsilon_{\rm initial}=0.04$. 
We consider $\tau_D=0$ms for (d) $\bar\varepsilon_{\rm initial}=0$, (e) 
$\bar\varepsilon_{\rm initial}=0.01$ and (f) $\bar\varepsilon_{\rm initial}=0.04$. 
$\tau_D=100$ms for (g) $\bar\varepsilon_{\rm initial}=0$, (h) 
$\bar\varepsilon_{\rm initial}=0.01$ and (i) $\bar\varepsilon_{\rm initial}=0.04$ 
and $\tau_D=1000$ms for (j) $\bar\varepsilon_{\rm initial}=0$, (k) 
$\bar\varepsilon_{\rm initial}=0.01$ and (l) $\bar\varepsilon_{\rm initial}=0.04$. 
Note that the synaptic weights $\varepsilon_{ij}$ (with $i,j=1,\ldots,100$) of the 
coupling matrices are encoded in grey scale in the colour bar.}
\label{fig7}
\end{center}
\end{figure*}

We performed a similar study for a HH neural network 
and analyzed how the connections evolve with the combined application of STP and STDP. In particular, 
we start with an all-to-all (globally connected) network of $N=100$ excitatory, HH neurons. The coupling weights 
vary in $[0,0.04]$ and $I_i$ is randomly distributed so that neural frequencies 
are in the range $[70,100]$Hz. We choose three initial coupling weight 
averages $\bar\varepsilon_{\rm initial}$. The coupling matrices for the connectivity strengths $\varepsilon_{ij}$ 
are shown in Fig. \ref{fig7} where $\bar\varepsilon_{\rm initial}=0$ in 
\ref{fig7}(a), $\bar\varepsilon_{\rm initial}=0.01$ in \ref{fig7}(b) and 
$\bar\varepsilon_{\rm initial}=0.04$ in \ref{fig7}(c). For all initial conditions studied,
we kept the standard 
deviation of $\bar\varepsilon_{\rm initial}$ fixed at $0.002$. In the 
coupling matrix $(\varepsilon_{ij})$, the presynaptic neurons $j$ and postsynaptic 
neurons $i$ are sorted and plotted in ascending frequency-order (i.e. from the 
smallest to the largest spike frequency). The final coupling matrices $\varepsilon_{ij}$ for 
$\tau_D=0$ms are shown in Fig. \ref{fig7}(d), (e) and (f) 
for $\bar\varepsilon_{\rm initial}=0$, $\bar\varepsilon_{\rm initial}=0.01$ and 
$\bar\varepsilon_{\rm initial}=0.04$, respectively. In all cases, the final 
coupling matrices have, predominantly, connections from faster to slower 
spiking neurons. This behaviour was also reported in Borges et al. 
\cite{borges17b} for a neural network with STDP. Figure \ref{fig7}(g), 
(h) and (i) present our results for 
$\bar\varepsilon_{\rm initial}=0$, $\bar\varepsilon_{\rm initial}=0.01$, 
and $\bar\varepsilon_{\rm initial}=0.04$, respectively, where $\tau_D=100$ms. 
Due to the effect of STP on the dynamics of the neurons in the network, we 
observe the formation of different modules of directly connected neurons. 
Again, the effect of STP leads to a decrease on the influence of the fastest 
neurons to the slowest ones, allowing for the formation of connections among 
those neurons with similar spike frequencies. The size of these modules 
increases according to the intensity of the initial coupling and their different 
sizes can be explained by the analysis made based on the results in Figs. 
\ref{fig5} and \ref{fig6}, where the coupling shortened the frequency differences,
leading to the formation of connections especially among neurons with close frequencies.
It also explains why the bigger-size modules are composed of the 
smallest-frequency spiking neurons. Increasing $\tau_D$ to 1000ms, leads to the 
disappearance of big modules and to the decrease of the number of coupled neurons,
as shown in Fig. \ref{fig7}(j), (k) and (l). Therefore, by 
varying the STP recovery time $\tau_D$, one can control the formation of modules 
in neural networks with STP and STDP.

To study further the observed modules, we consider connections with
$\varepsilon_{ij}>0.002$ (i.e. with connectivity strength bigger than
$5\%$ of the maximal coupling strength). Thus, weaker connections are
not considered in the resulting network analysis. This procedure avoids
measurement errors that might be caused by connections whose weights
fluctuate over time closely to zero.

To evaluate how modular structures evolve over time, we compute the modularity
$Q$ by using the Louvain method \cite{Blondel}. $Q$ is measured across network
partitions in densely connected communities. In particular, the modularity
assumes values in the range $[-1,1]$ comparing the density of connections
within communities with the density among communities. The best network partition
in modules is one that maximizes modularity. $Q$ is defined as \cite{Newman2004}
\begin{eqnarray}
 Q &=& \frac{1}{W}\sum_{i}^{N}\sum_{j}^{N}\left ( \varepsilon_{i,j} - \frac{\omega_i\omega_j}{W}  \right )\delta(c_i,c_j) \label{Q},
\end{eqnarray}
where $\omega_i=\sum_{j}^{N}\varepsilon_{ij}$ represents the sum of the connection
weights received by node $i$ and $W=\sum_{i}^{N}\sum_{j}^{N}\varepsilon_{ij}$ is
the sum of all weights in the coupling matrix. The term $c_i$ represents the
community that neuron $i$ has been allocated to and $\delta(c_i,c_j)$ is given by
\begin{eqnarray}
\delta(c_i,c_j) &=& \left\{\begin{matrix}
1,\mbox{ if }c_i=c_j,\\ 
0,\mbox{ otherwise}.
\end{matrix}\right.
\end{eqnarray}
The Louvain \cite{Blondel} method is defined in two steps. At first, each node
in the network is considered as a community in itself, and thus initially, there will
be as many communities as nodes in the network. At this stage, each node
$i$ is reassigned to the community of each of its neighbors $j$,
then $i$ will be permanently fixed in the community that promotes the largest
gain in modularity $Q$ (positive gain). This process is applied repeatedly to
all network nodes until there are no more gains in $Q$. The second step amounts
to taking the defined communities at the end of the first step and consider them as
the nodes of a new network. The weight of the connections between these new nodes
is given by the sum of the weights of the connections between the nodes present in
the communities (defined in the first step). Once the new network is computed,
the first step can be applied again to its nodes. This process occurs repeatedly
until no further changes to $Q$ occur and a maximum value is thus obtained.

\begin{figure}[htbp]
\begin{center}
\centering\includegraphics[width=0.4\textwidth]{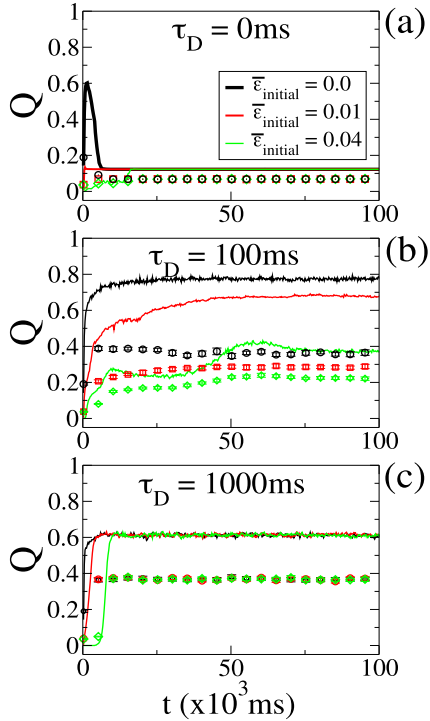}
\caption{Time evolution of the modularity $Q$ for (a) $\tau_D=0$ms, (b) $\tau_D=100$ms and
(c) $\tau_D=1000$ms. $Q$ was calculated over the time evolution of the coupling matrix and
the coloured curves represent the network average initial coupling: the black curve is for
${\bar \varepsilon}_{\rm initial}=0$, the red for ${\bar \varepsilon}_{\rm initial}=0.01$
and the green for ${\bar \varepsilon}_{\rm initial}=0.04$. The colored symbols represent
the average $Q$ for random networks computed by rewiring the connections in the corresponding
original networks. For example, the black symbols represent the average $Q$ calculated over
20 random networks generated from rewiring randomly the connections in the original networks (black curve).}
\label{fig8}
\end{center}
\end{figure}

Figure \ref{fig8} shows the time evolution of the modularity $Q$ for the three
recovery times, $\tau_D=0$ms, $\tau_D=100$ms and $\tau_D=1000$ms. The colours
represent the initial connectivity strengths in the coupling matrix: the black
curve is for ${\bar\varepsilon}_{\rm initial}=0$, the red curve for
${\bar\varepsilon}_{\rm initial}=0.01$ and the green curve for ${\bar\varepsilon}_{\rm initial}=0.04$.
The coloured symbols represent the average $Q$ values calculated over 20 random networks obtained by
rewiring all connections of the original networks. We did this to compare our results with those
obtained for random networks with the same connections and number of nodes (which we call
random variants). Unless stated otherwise, the symbols in Fig. \ref{fig8} represent the
measurements taken on these random variants. In Fig. \ref{fig8}(a), we observe that for
$\tau_D=0$ms (instantaneous recovery), $Q$ is very low for all ${\bar\varepsilon}_{\rm initial}$
connectivity strengths, remaining constant after a small transient. These results confirm those
in Fig. \ref{fig7}(d), (e) and (f). In Fig. \ref{fig8}(b) for $\tau_D=100$ms, the modularity $Q$
takes its largest value for ${\bar\varepsilon}_{\rm initial}=0$, confirming what we observed by
comparing the number of modules in Fig. \ref{fig7}(g), (h) and (i). It is also apparent that $Q$
stabilizes for simulation times $t$ greater than $80000$ms as there are no changes in the modular
structures occurring in the networks anymore. Interestingly enough, Fig. \ref{fig8}(c) shows that
for $\tau_D=1000$ms, the modularity converges to the same value (i.e. $Q\approx 0.6$) for all
${\bar\varepsilon}_{\rm initial}$ coupling strengths, again in accordance with the results in
Fig. \ref{fig7}(j), (k), (l), which show networks with similar configurations regardless of the
initial coupling. Finally, in all cases considered, we find that the modularity of the networks
is bigger than their random variants.

Next, we compute other quantities that characterize the structure of the networks considered
previously, such as the mean path-length, clustering coefficient and assortativity.

In particular, a path is defined as the route that passes through network connections
connecting two nodes $i$ and $j$. The path with the shortest number of connections is
called the shortest path $l_{ij}$ \cite{barabasi}. The networks we consider here are
directed, so $l_{ij}$ is not necessarily equal to $l_{ji}$. In general, the shortest
average path length is given by
\begin{eqnarray}
{\bar L} &=& \frac{1}{N(N-1)}\sum_{i}^{N}\sum_{j\neq i}^{N}l_{ij},\label{L}
\end{eqnarray}
where $N$ is the number of nodes in the network. In our approach, we calculated ${\bar L}$
via a breadth-first search approach \cite{barabasi} and do not consider the weights of the
connections. If there is no possible directed path $l_{ij}$ between nodes $i$ and $j$, then
it is not considered in the calculations in Eq. \eqref{L}.

\begin{figure}[htbp]
\begin{center}
\centering\includegraphics[width=0.3\textwidth]{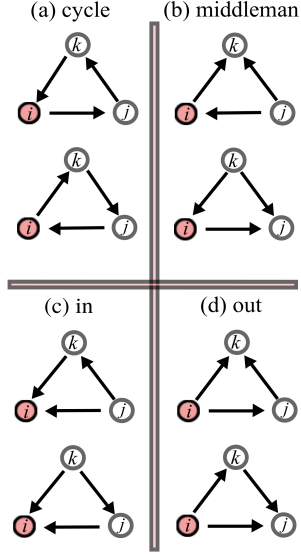}
\caption{The 8 distinct triangular motifs in directed networks. The motifs are considered
with respect to node $i$ depicted in red and are split into the cycle, middleman, in and out motifs.}
\label{fig9}
\end{center}
\end{figure}

The second computed quantity is the clustering coefficient ${\rm CC}_i$, which measures the degree
to which the neighbors of a node $i$ are connected to each other and varies in $[0,1]$. It is computed
by considering the number of triangular motifs made by node $i$ and its neighbors compared to all
possible triangular motifs of that node \cite{barabasi}. For directed networks, given 3 connected nodes $i$, $j$ and $h$,
there are 8 distinct triangular motifs, shown in Fig. \ref{fig9}. These motifs are further organised
into 4 groups when considering node $i$ as the reference node: Figure \ref{fig9}(a) shows a ``cycle''
motif, Fig. \ref{fig9}(b) a ``middleman'' motif, Fig. \ref{fig9}(c) an ``in'' motif and
Fig. \ref{fig9}(d) an ``out'' motif. For each motif, we calculate
${\rm CC}_i^{(\mbox{cyc},\mbox{mid},\mbox{in},\mbox{out})}$ relative to node $i$, as shown in\cite{Fagiolo}.
For directed and weighted networks, there are 4 types of clustering coefficients
\begin{eqnarray}
{\rm CC}_i^{\mbox{cyc}} &=& \frac{ \frac{1}{2} \sum_{j}^{N}\sum_{h}^{N}\left [ \varepsilon_{ij}^{1/3}\varepsilon_{jh}^{1/3}\varepsilon_{hi}^{1/3}+\varepsilon_{ih}^{1/3}\varepsilon_{hj}^{1/3}\varepsilon_{ji}^{1/3}  \right ]}{d_i^{\mbox{in}}d_i^{\mbox{out}} - d_i^{\leftrightarrow}},\nonumber
\end{eqnarray}
\begin{eqnarray}
{\rm CC}_i^{\mbox{mid}} &=& \frac{ \frac{1}{2} \sum_{j}^{N}\sum_{h}^{N}\left [ \varepsilon_{ih}^{1/3}\varepsilon_{jh}^{1/3}\varepsilon_{ji}^{1/3}+\varepsilon_{ij}^{1/3}\varepsilon_{hi}^{1/3}\varepsilon_{hj}^{1/3}  \right ]}{d_i^{\mbox{in}}d_i^{\mbox{out}} - d_i^{\leftrightarrow}},\nonumber
\end{eqnarray}
\begin{eqnarray}
{\rm CC}_i^{\mbox{in}} &=& \frac{ \frac{1}{2} \sum_{j}^{N}\sum_{h}^{N}\left [ \varepsilon_{ij}^{1/3}\varepsilon_{ih}^{1/3}\varepsilon_{jh}^{1/3}+\varepsilon_{ij}^{1/3}\varepsilon_{ih}^{1/3}\varepsilon_{hj}^{1/3}  \right ]}{d_i^{\mbox{in}}(d_i^{\mbox{in}} - 1)},\nonumber
\end{eqnarray}
\begin{eqnarray}
{\rm CC}_i^{\mbox{out}} &=& \frac{ \frac{1}{2} \sum_{j}^{N}\sum_{h}^{N}\left [ \varepsilon_{hi}^{1/3}\varepsilon_{ji}^{1/3}\varepsilon_{jh}^{1/3}+\varepsilon_{hi}^{1/3}\varepsilon_{ji}^{1/3}\varepsilon_{hj}^{1/3}  \right ]}{d_i^{\mbox{out}}(d_i^{\mbox{out}} - 1)},\label{CCi}
\end{eqnarray}
where $d_i^{\mbox{in}}=\sum_{j}^{N}a_{ij}$ is the in-degree of node $i$ and
$d_i^{\mbox{out}}=\sum_{j}^{N}a_{ji}$ its out-degree. The term
$d_i^{\leftrightarrow}=\sum_{j}^{N}a_{ij}a_{ji}$ represents the number of bilateral
connections between node $i$ and its neighbors. For the calculation of $d_i^{\mbox{in}}$,
$d_i^{\mbox{out}}$ and $d_i^{\leftrightarrow}$, the coupling weigths are not considered, only the
number of connections, that is $a_{ij}=1$ if $\varepsilon_{ij}>0.002$, otherwise $a_{ij}=0$.

Consequently, the clustering coefficient of the network, ${\rm CC}^*$, is calculated by
averaging ${\rm CC}_i^*$ over all $N$ nodes in the network
\begin{eqnarray}
{\rm CC}^* &=&  \frac{1}{N}\sum_{i}^{N}{\rm CC}_i^*,\label{CC}
\end{eqnarray}
where $*$ stands for either of the cyc, mid, in or out motifs.
 
The last quantity in our study is assortativity, which is the correlation coefficient
(i.e. the Pearson correlation \cite{Newman2002}) between the degrees of nodes on two opposite ends
of a connection in a network, for all connections in the network. This correlation varies in $[-1,1]$
and, is positive in assortative networks and negative in disassortative networks. Since our networks
are directed and weighted, we use the four directed assortativity measures defined in \cite{Foster}.

\begin{figure}[htbp]
\begin{center}
\centering\includegraphics[width=0.4\textwidth]{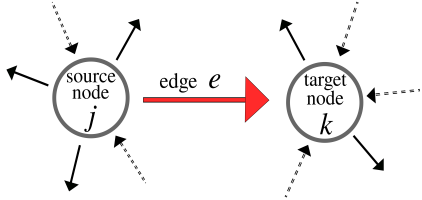}
\caption{Representative illustration of an edge $e$ and, the source $j$ and target $k$ nodes
considered in Eq. \eqref{assor}.}
\label{fig10}
\end{center}
\end{figure}

In particular, let $a$ and $b$ be indices representing the type of degree (i.e. in- and out-degree)
and $j^a_i$ and $k^b_i$ the $a$- and $b$-degree from source node $j$ to target node $k$ of an edge
$e$. Figure \ref{fig10} shows an illustrative representation of edge $e$ and their respective source
nodes $j$ and target nodes $k$. Foster et al. \cite{Foster} defined the assortativity measure $r(a,b)$ as
\begin{eqnarray}
r(a,b) &=& \frac{N_e^{-1}\sum_{e}^{N_e}\left [ (j_e^a - {\bar j^a})(k_e^b - {\bar k^b}) \right ]}{\sigma^a\sigma^b}, \label{assor}
\end{eqnarray}
where $N_e$ is the number of edges in the network, ${\bar j^a}=N_e^{-1}\sum_{e}^{N_e}j_e^a$ the
average in- and out-degrees of the source node and,
$\sigma^a=\sqrt{N_e^{-1}\sum_{e}^{N_e^{-1}}(j_e^a-{\bar j^a})^2}$ its deviation, calculated for
all edges. The equations for ${\bar k^b}$ and $\sigma^b$ are similarly defined. In our approach,
instead of in- and out-degrees, we use the in- and out-strength from the source and target nodes,
given by the sum of their in- and out-coupling weights. A positive assortativity coefficient
indicates that nodes tend to connect to other nodes with the same or similar strength.
 
\begin{figure*}[htbp]
\begin{center}
\centering\includegraphics[width=1.0\textwidth]{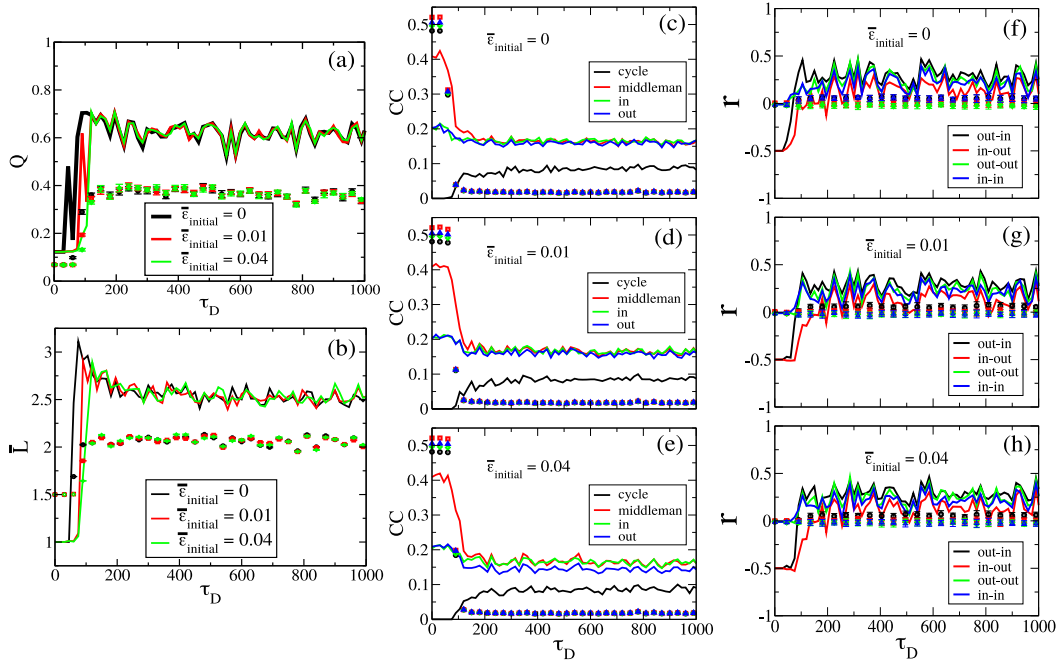}
\caption{Structural properties of the final coupling matrix configuration (coloured curves) and their respective
random variants (averaged over 20 random networks, coloured symbols) as a function of the recovery time $\tau_D$.
Panels (a) and (b) show the modularity $Q$ and average shortest path length ${\bar L}$ for three different
average initial couplings: ${\bar \varepsilon}_{\rm initial}=0$ in black, ${\bar \varepsilon}_{\rm initial}=0.01$
in red and ${\bar \varepsilon}_{\rm initial}=0.04$ in green. In (c), (d) and (e), we show all clustering
coefficients for ${\bar \varepsilon}_{\rm initial}=0$, ${\bar \varepsilon}_{\rm initial}=0.01$ and
${\bar \varepsilon}_{\rm initial}=0.04$, respectively. Panels (f), (g) and (h) show the assortativity for
${\bar \varepsilon}_{\rm initial}=0$, ${\bar \varepsilon}_{\rm initial}=0.01$ and ${\bar \varepsilon}_{\rm initial}=0.04$, respectively.}
\label{fig11}
\end{center}
\end{figure*}

Figure \ref{fig11} shows the results of the computations of the quantities discussed previously
as a function of $\tau_D$ in $[0,1000]$ms. In all computations, we consider the final coupling
matrix obtained after $t=200000$ms of simulated data. Figure \ref{fig11}(a) shows the modularity
$Q$ for 3 initial couplings: ${\bar \varepsilon}_{\rm initial}=0$ in black,
${\bar \varepsilon}_{\rm initial}=0.01$ in red and, ${\bar \varepsilon}_{\rm initial}=0.04$ in
green and their random variants given by different symbols with their respective colors. We
note that for $\tau_D<30$ms (approximately) and for all 3 ${\bar \varepsilon}_{\rm initial}$
values, $Q$ is low. For $30\mbox{ms}<\tau_D<120$ms (approximately), $Q$ depends on
${\bar \varepsilon}_{\rm initial}$. In particular, the lower the initial average coupling
${\bar \varepsilon}_{\rm initial}$, the more modular the final configuration of the network.
For $\tau_D>120$ms (approximately), there is no dependence on ${\bar \varepsilon}_{\rm initial}$
and the resulting networks exhibit similar modularity values. In all cases of
${\bar \varepsilon}_{\rm initial}$, $Q$ is higher than that calculated for their random variants,
which shows that the resulting networks do not have characteristics of random networks.

Figure \ref{fig11}(b) shows the results for the average path length ${\bar L}$ for the same three
${\bar \varepsilon}_{\rm initial}$ values. For $\tau_D<30$ms, the average path ${\bar L}=1$, in
accordance with Fig. \ref{fig7}(d), (e) and (f), where we observe a network where all nodes are
connected to all other nodes with unidirectional connections. For $30\mbox{ms}<\tau_D<120$ms,
the average path length ${\bar L}$ is bigger for weaker initial couplings. This is because weak
initial couplings result in the formation of a larger number of modular structures (as seen in
Fig. \ref{fig11}(a)), which results in a greater path for one node to access other nodes in different
modules, possibly crossing through other modules. For $\tau_D>120$ms, we see that
${\bar L}\approx 2.5$ for all 3 ${\bar \varepsilon}_{\rm initial}$ values. In contrast, the average
path lengths ${\bar L}$ for the random variants are smaller than those for the original networks.

Figure \ref{fig11}(c), (d) and (e) present the 4 types of clustering coefficients for
${\bar \varepsilon}_{\rm initial}=0$, ${\bar \varepsilon}_{\rm initial}=0.01$ and
${\bar \varepsilon}_{\rm initial}=0.04$. The black curve represents ${\rm CC}^{\mbox{cyc}}$, the
red ${\rm CC}^{\mbox{mid}}$, the green ${\rm CC}^{\mbox{in}}$ and the blue ${\rm CC}^{\mbox{out}}$.
In all cases, the results are similar. It is worth it to note that for $\tau_D<120$ms,
${\rm CC}^{\mbox{mid}}$ has the highest value and ${\rm CC}^{\mbox{cyc}}$ the lowest. These results
build on what we have already observed: the action of STDP promotes connections from faster to
slower spiking neurons and do not permit cyclic connections (see Fig. \ref{fig9}(a)).
As $\tau_D$ increases, ${\rm CC}^{\mbox{mid}}$ converges to 
${\rm CC}^{\mbox{in}},{\rm CC}^{\mbox{out}}\approx 0.15$ and ${\rm CC}^{\mbox{cyc}}$ increases
to $\approx 0.09$. For the random variants of the networks (coloured symbols), we see that for
$\tau_D<120$ms, all clustering coefficients are similar and larger than those for the original
networks. This is because for such $\tau_D$ values, the networks from which they were generated
(Fig. \ref{fig7}(d), (e), (f)) are densely connected and the high number of connections allows
for the formation of triangular motifs without any of the 4 types occuring preferentially. For
$\tau_D>120$, all random networks have their clustering coefficients fluctuate near zero. In
this case, the generating networks are more sparse (as in Fig. \ref{fig7}(g), (h), (i)), which
makes their random variants have low probability in forming triangular motifs. We note here that these
clustering coefficients are not high enough to claim with certainty that the networks
have a small-world topology.

The right column of plots in Fig. \ref{fig11} shows the 4 assortativity measures, $r$, computed
for the considered networks with ${\bar \varepsilon}_{\rm initial}=0$,
${\bar \varepsilon}_{\rm initial}=0.01$ and ${\bar \varepsilon}_{\rm initial}=0.04$. The black
curves represent the out-strength/in-strength correlation $r(\mbox{out},\mbox{in})$ between
the source and target nodes, the red curve the in-strength/out-strength correlation
$r(\mbox{in},\mbox{out})$, the green curve the out-strength/out-strength correlation
$r(\mbox{out},\mbox{out})$ and the blue curve, the in-strength/in-strength correlation
$r(\mbox{in},\mbox{in})$. In Fig. \ref{fig11}(f), (g), (h) and for the three
${\bar \varepsilon_{\rm{initial}}}$ values, we find that for $\tau_D <100$ms, the networks are
disassortative as $r(\mbox{out}, \mbox{in})$ and $r(\mbox{in},\mbox{out})$ are negative and,
at the same time, neither assortative nor disassortative as $r(\mbox{out},\mbox{out})$ and
$r(\mbox{in}, \mbox{in})$ are approximately equal to 0. As $\tau_D$ increases, all correlations
grow being mostly positive with $r(\mbox{out}, \mbox{in})$ being the largest. This shows a
greater correlation in the network of high out-strength nodes to connect with nodes with
high in-strength and corroborates the results in Fig. \ref{fig7} where connections occur
preferably from the faster neurons (which have a high out-strength) to the slower ones
(with high in-strength). For the random variants of these networks, all correlations are
close to zero for the entire $\tau_D$ range, a result completely different to those for the
original networks. We thus conclude that the original networks are far from being purely random
networks, exhibiting a type of preferential attachment in their connectivities.

In the analysis of the structural properties of the networks, the topology may vary greatly
depending on $\tau_D$. In all cases studied, the networks were different from random or
small-world networks, since the average path lengths are bigger than in random networks and
their clustering coefficients are small than in random networks. This corroborates further
the results obtained previously here that there is a complex configuration where modules are
formed following a kind of preferential attachment process.

\section{CONCLUSIONS}

In this paper, we studied the effects of plasticity (STP and STDP) on networks of excitatory
coupled Hodgkin-Huxley neurons. Neural plasticity is responsible for alterations in the organisation
and  structure of the brain, and both play an important role in synaptic weights. Besides, STDP has
a longer time scale than STP, so it affects differently the structure and function in brain networks.

We started analysing the effect of STP in a pair of neurons that are initially, either uncoupled or
bidirectionally coupled. For initially uncoupled neurons, the action of STP and STDP promotes directed
connections among neurons with small spike frequency differences, from the faster to the slower spiking
neurons. The increase of the recovery-time shortened the interval of frequency differences where
connections are formed. When neurons are initially coupled, their frequency difference is smaller
and increases the size of the area of directed connections. We found that  STP induces uncoupling,
depending on the recovery time: the bigger the recovery time, the smaller the interval of frequency
difference that allows for the formation of connections.

Next, we build a neural network with an all-to-all topology. Considering only STDP, the
coupling matrix exhibits directed connections from neurons with high to neurons  with low spike
frequencies. We have shown that due to STP, neural networks equipped with STDP facilitate the
formation of synapses among neurons with similar spike frequencies only and that modular neural
networks can emerge as a direct result of the combined effect of STP and STDP, a phenomenal structure
also depicted by neurophysiological and experimental studies. However, by increasing the STP recovery
time, the number of connections decreased and as a consequence, the modules disappeared. That is
actually a way to control the modular organization in neural networks. The structure of these modular
networks is complex, unlike those in random or small-world networks, resembling more to networks with
preferential attachment properties.

In future, we plan to study neural networks with greater diversity in chemical synapses, addressing
other STP and STDP rules related to synapses between excitatory-inhibitory and inhibitory-inhibitory
neurons \cite{caporale2008}. Finally, another interesting aspect of our work would be the introduction
of time delay in the synaptic transmission to study how it affects the evolution of the couplings and
the modular properties of neural networks.

\begin{acknowledgments}
We wish to acknowledge Dr. Serhiy Yanchuk for the valuable suggestions and remarks. Also the
support offered by the International Visiting Fellowships Scheme of
the University of Essex, CNPq, CAPES, Funda\c c\~ao Arauc\'aria and S\~ao Paulo Research
Foundation (processes FAPESP 2011/19296-1, 2015/07311-7,  2016/23398-8, 2017/13502-5, 2017/18977-1, 2017/20920-8).
The research was also supported by the 2015/50122-0 S\~ao Paulo Research Foundation (FAPESP)
and DFG-IRTG 1740/2 grants.
\end{acknowledgments}



\begin{thebibliography}{00}
\bibitem{tewari16}
S. G. Tewari, M. K. Gottipati, and V. Parpura, ``Mathematical modeling in
neuroscience: Neural activity and its modulation by astrocytes", Front.
Integr. Neurosci. {\bf 10}, 3 (2016).
\bibitem{lapicque07}
L. Lapicque, ``Recherches quantitatives sur l'excitation electrique des nerfs
trait\'ee comme une polarization", J. Physiol. Pathol. Gen. {\bf 9}, 620-635
(1907).
\bibitem{hodgkin52}
A. L. Hodgkin and A. F. Huxley, ``A quantitative description of membrane
current and its application to conduction and excitation in nerve", J.
Physiol. {\bf 117}, 500-544 (1952).
\bibitem{zhu18}
Z. Zhu, R. Wang, and F. Zhu, ``The energy coding of a structural neural network
based on the Hodgkin-Huxley model", Front. Neurosci. {\bf 12}, 122 (2018).
\bibitem{Protacheviczetal2019}
P. R. Protachevicz, F. Borges, E. L. Lameu, P. Ji, K. C. Iarosz, A. H. Kihara, 
I. L. Caldas, J. D. Szezech, M. S. Baptista, E. N. Macau, Ch. G. Antonopoulos, 
A. M. Batista, and J. Kurths, ``Bistable firing pattern in a neural network model'', 
Front. Comput. Neurosc. {\bf 13}, 19 (2019).
\bibitem{Viana14}
R.L. Viana, F.S. Borges, K.C. Iarosz, A.M. Batista, S.R. Lopes, and I.L. Caldas,
``Dynamic range in a neuron network with electrical and chemical synapses",
Commun. Nonlinear Sci. Numer. Simul. {\bf 19}, 164–172 (2014).
\bibitem{batista14}
C. A. S. Batista, R. L. Viana, S. R. Lopes, and A. M. Batista, ``Dynamic range
in small-world networks of Hodgkin-Huxley neurons with chemical synapses",
Physica A {bf 410}, 628-640 (2014).
\bibitem{Borges2015}
F.S. Borges, E.L. Lameu, A.M. Batista, K.C. Iarosz, M.S. Baptista, and R.L. Viana, 
``Complementary action of chemical and electrical synapses to perception",
Physica A {\bf 430}, 236–241 (2015).
\bibitem{Antonopoulos2016}
Ch. Antonopoulos, ``Dynamic range in the C.elegans brain network'', Chaos, {\bf 26}, 1, 1054-1500 (2016).

\bibitem{Hizanidisetal2016} 
J. Hizanidis, N. E.  Kouvaris, G. Zamora-L\'opez, 
A. D\'iaz-Guilera, and Ch. G. Antonopoulos, ``Chimera-like states in modular 
neural networks'', Scientific Reports {\bf 6} 19845 (2016).
\bibitem{Borges2017}
F. S. Borges,  P. R. Protachevicz, E. L. Lameu, R. C. Boneti, K. C. Iarosz, I. L. Caldas, 
M. S. Baptista, and A. M. Batista, ``Synchronised fire patterns in a random network of 
adaptive exponential integrate-and-fire neuron model", Neural Networks, {\bf 90}, 1-7 (2017).
\bibitem{protachevicz18}
P. R. Protachevicz, R. R. Borges, F. S. Borges, K. C. Iarosz, I. L. Caldas, E.
L. Lameu, E. E. N. Macau, R. L. Viana, I. M. Sokolov, F. A. S. Ferrari, J.
Kurths, A. M. Batista, C.-Y. Lo, Y. He, and C.-P. Lin, ``Synchronous behaviour
in network model based on human cortico-cortical connections", Physiol. Meas.
{\bf 39}, 074006 (2018).
\bibitem{lameu18a}
E. L. Lameu, S. Yanchuk, E. E. N. Macau, F. S. Borges, K. C. Iarosz, I. L.
Caldas, P. R. Protachevicz, R. R. Borges, R. L. Viana, J. D. Szezech Jr., A. M.
Batista, and J. Kurths, ``Recurrence quantification analysis for the
identification of burst phase synchronization", Chaos {\bf 28}, 085701 (2018).
\bibitem{Antonopoulosetal2019}
Antonopoulos Ch. G., Martinez-Bianco E. and Baptista M., ``Evaluating performance of 
neural codes in model neural communication networks", Neural Networks {\bf 109}, 90-102 (2019).
\bibitem{Antonopoulosetal2015}
Ch. G. Antonopoulos,  S. Srivastava, S. E. D. S. Pinto, and 
M. S. Baptista, ``Do brain networks evolve by maximizing their information flow 
capacity?'', PLOS Computational Biology {\bf 11}, 8, e1004372-e1004372 (2015).
\bibitem{Antonopoulosetal2016}
Ch. G. Antonopoulos, A. S. Fokas, and T. C. Bountis, ``Dynamical complexity in the 
C.elegans neural network'', European Physical Journal: Special Topics {\bf 225},6-7, 1255-1269 (2016).
\bibitem{borges18}
F. S. Borges, E. L. Lameu, K. C. Iarosz, P. R. Protachevicz, I. L. Caldas, R. L.
Viana, E. E. N. Macau, A. M. Batista, and M. S. Baptista, ``Inference of
topology and the nature of synapses, and the flow of information in neural
networks", Phys. Rev. E {\bf 97}, 022303 (2018).
\bibitem{romani15}
S. Romani and M. Tsodyks, ``Short-term plasticity based network model of place
cells dynamics", Hippocampus {\bf 25}, 94-105 (2015).
\bibitem{zenke15}
F. Zenke, E. J. Agnes, and W. Gerstner, ``Diverse synaptic plasticity
mechanisms orchestrated to form and retrieve memories in spiking neural
networks", Nat. Commun. {\bf 6}, 6922 (2015).
\bibitem{borges17b}
R. R. Borges, F. S. Borges, E. L. Lameu, A. M. Batista, K. C. Iarosz, I. L.
Caldas, C. G. Antonopoulos, and M. S. Baptista, ``Spike-timing-dependent
plasticity induces non-trivial topology in the brain", Neural Networks, {\bf 88},
58-64 (2017).
\bibitem{burke06}
S. N. Burke and C. A. Barnes, ``Neural plasticity in the ageing brain", Nature
Rev. Neurosci. {\bf 7}, 30-40 (2006).
\bibitem{berlucchi09}
G. Berlucchi and H. A. Buchtel, ``Neural plasticity: historical roots and
evolution of meaning", Exp. Brain Res. {\bf 192}, 307-319 (2009).
\bibitem{james90}
W. James, ``The principles of psychology", Chapter IV, Habits (1890).
\bibitem{stahnisch02}
F. W. Stahnisch and R. Nitsch, ``Santiago Ram\'on y Cajal's concept of neural
plasticity: the ambiguity lives on", Trends Neurosci. {\bf 25}, 589-591 (2002).
\bibitem{lashley24}
K. S. Lashley, ``Studies of cerebral function in learning. VI. The theory that
synaptic resistance is reduced by the passage of the nerve impulse", Psychol.
Rev. {\bf 31}, 369-375 (1924).
\bibitem{konorski48}
J. Konorski, ``Conditioned reflexes and neuron organization", Cambridge
University Press, Cambridge (1948).
\bibitem{hebb49}
D. O. Hebb, ``The organization of behaviour. A neuropsychological theory",
Wiley, New York (1949).
\bibitem{bennett64}
E. L. Bennett, M. C. Diamond, D. Krech, and M. R. Rosenzweig, ``Chemical and
anatomical plasticity of the brain", Science {\bf 146}, 610-619 (1964).
\bibitem{lameu18b}
E. L. Lameu, E. E. N. Macau, F. S. Borges, K. C. Iarosz, I. L. Caldas, R. R.
Borges, P. R. Protachevicz, R. L. Viana, and A. M. Batista, ``Alterations in
brain connectivity due to plasticity and synaptic delay", Eur. Phys. J. Spec.
Top. {\bf 227}, 673-682 (2018).
\bibitem{rangaraju19}
V. Rangaraju, M. Lauterbach, and E. M. Schuman, ``Spatially stable
mitochondrial compartments fuel local translation during plasticity", Cell
{\bf 176}, 73-84 (2019).
\bibitem{abbott00}
L. F. Abbott and S. B. Nelson, ``Synaptic plasticity: taming the beast",
Nat. Neurosci. {\bf 3}, 1178-1183 (2000).
\bibitem{borges17a}
R. R. Borges, F. S. Borges, E. L. Lameu, P. R. Protachevicz, K. C. Iarosz, I. L.
Caldas, R. L. Viana, E. E. N. Macau, M. S. Baptista, C. Grebogi, and A. M.
Batista, ``Synaptic plasticity and spike synchronization in neural
networks", Braz. J. Phys. {\bf 47}, 678-688 (2017).
\bibitem{mcdonnell17}
M. D. McDonnell and B. P. Graham, ``Phase changes in neural postsynaptic
spiking due to short term plasticity", PLoS Comput. Biol. {\bf 13}, e1005634
(2017).
\bibitem{tass18}
M. M. Asl, A. Valizadeh, and P. A. Tass, ``Delay-Induced Multistability
and Loop Formation in Neuronal Networks with Spike-Timing-Dependent Plasticity'',
Sci. Reports {\bf 8}, 12068 (2018).
\bibitem{markram12}
H. Markram, W. Gerstner, and P. J. Sj\"ostr\"om, ``Spike-timing-dependent
plasticity: a comprehensi\-ve overview", Front. Synaptic Neurosci. {\bf 4},
1-3 (2012).
\bibitem{borges16}
R. R. Borges, F. S. Borges, E. L. Lameu, A. M. Batista, K. C. Iarosz, I. L.
Caldas, R. L. Viana, and M. A. F. Sanju\'an, ``Effects of the spike
timing-dependent plasticity on the synchronization in a random Hodgkin-Huxley
neural network", Commun. Nonlinear Sci. Numer. Simulat. {\bf 34}, 12-22
(2016).
\bibitem{clopath10}
C. Clopath , L. B\"using, E. Vasilaki, W. Gerstner, ``Connectivity reflects coding: a
model of voltage-based STDP with homeostasis", Nat Neurosci {\bf 13}, 344–352 (2010).
\bibitem{tass12}
P. A. Tass, O. V. Popovych, ``Unlearning tinnitus-related cerebral synchrony with
acoustic coordinated reset stimulation: theoretical concept and modelling", Biol.
Cybern. {\bf 106}, 27–36 (2012).
\bibitem{popovych13}
O. V. Popovych, S. Yanchuk, and P. A. Tass, ``Self-organized noise resistance
of oscillatory neural networks with spike-timing-dependent plasticity", Sci.
Rep. {\bf 3}, 2926 (2013).
\bibitem{lucken16}
L. Lücken, O. V. Popovych, P. Tass, and S. Yanchuk, ``Noise-enhanced coupling between
two oscillators with long-term plasticity", Phys. Rev. E {\bf 93}, 32210 (2016).
\bibitem{gestner96}
W. Gerstner, R. Kempter, J. L. van Hemmen, H. Wagner, ``A neuronal learning rule
for sub-millisecond temporal coding", Nature {\bf 383}, 76–78 (1996).
\bibitem{stevens95}
C. F. Stevens and Y. Wang, ``Facilitation and depression at single central synapses", Neuron {\bf 14}, 795-802 (1995).
\bibitem{abbott97}
L. F. Abbott, J. A. Varela, K. Sen, and S. B. Nelson, ``Synaptic
depression and cortical gain control", Science {\bf 275}, 220-224 (1997).
\bibitem{zucker02}
R. S. Zucker and W. G. Regehr, ``Short-term synaptic plasticity", Annu. Rev.
Physiol. {\bf 64}, 355-405 (2002).
\bibitem{hennig13}
M. H. Hennig, ``Theoretical model of synaptic short term plasticity", Front.
Comput. Neurosci. {\bf 7}, 45 (2013).
\bibitem{itskov11}
V. Itskov, D. Hansel, and M. Tsodyks, ``Short-term facilitation may stabilize
parametric working memory trace", Front. Comput. Neurosci. {\bf 5}, 40 (2011).
\bibitem{york09}
L. C York and M. C. W. van Rossum, ``Recurrent networks with short term
synaptic depression", J. of Comp. Neuroscience, {\bf 27}, 607-620 (2009).
\bibitem{bi98}
G. Q. Bi and M. M. Poo, ``Synaptic modifications in cultured hippocampal
neurons: Dependence on spike timing, synaptic strength, and postsynaptic cell
type", J. Neurosci. {\bf 18}, 10464-10472 (1998).
\bibitem{bi01}
G. Q. Bi and M. M. Poo, ``Synaptic modification by correlated activity: Hebb's
postulate revisited", Annu. Rev. Neurosci. {\bf 24}, 139-166 (2001).
\bibitem{liu03}
S.-C. Liu, ``Analog VLSI circuits for short-term dynamic synapses", EURASIP
J. Appl. Signal Process. {\bf 7}, 620-628 (2003).
\bibitem{zucker89}
R. S. Zucher, ``Short-term synaptic plasticity", Ann. Rev. Neurosci. {\bf 12},
13-31 (1989).
\bibitem{bifone16}
C. Nicolini and A. Bifone, ``Modular structure of brain
functional networks: breaking the resolution limit by Surprise", 
Sci. Reports {\bf 6}, 19250 (2016).
\bibitem{betzel16}
O. Sporns, R. F. Betzel, ``Modular Brain Networks", Annu. Rev. Psychol. 
{\bf 67}, 613-640 (2016).

\bibitem{Shannon}
C. E. Shannon, ``A mathematical theory of communication", Bell Syst.
Tech. J. {\bf 27}, 379-423, 623-656 (1948).

\bibitem{Kullback}
S. Kullback, ``Information Theory and Statistics", Wiley, New York (1959).

\bibitem{Dobrushin}
R. L. Dobrushin, ``A general formulation of the fundamental theorem of Shannon in the theory of information'', Usp. Mat. Nauk. {\bf 14}, 6, 3-104 (1959).

\bibitem{caporale2008}
N. Caporale and Y. Dan, ``Spike Timing-Dependent Plasticity: A Hebbian
Learning Rule'', Annu. Rev. Neurosci. {\bf 31}, 25-46 (2008).

\bibitem{Blondel}
V. D. Blondel, J. Guillaume, R. Lambiotte, and E. Lefebvre, ``Fast unfolding
of communities in large networks", J. Stat. Mech. {\bf 08}, P10008 (2008).

\bibitem{Newman2004}
M. E. J. Newman, ``Analysis of weighted networks", Phys. Rev. E {\bf 70}, 056131 (2004).

\bibitem{barabasi}
A. Barab\'asi, and M. P\'osfai, ``Network Science", Cambridge University Press, Cambridge (2016).

\bibitem{Fagiolo}
G. Fagiolo, ``Clustering in complex directed networks", Phys. Rev. E {\bf 76}, 026107 (2007).

\bibitem{Newman2002}
M. E. J. Newman, ``Assortative mixing in networks", Phys. Rev. Letters {\bf 89}, 208701 (2002)

\bibitem{Foster}
J. G. Foster, D. V. Foster, P. Grassberger and M. Paczuski, ``Edge direction and the structure
of networks", PNAS {\bf 107}, 10815-10820 (2010).

\end{thebibliography}
\end{document}